\pdfoutput=1

\documentclass[aps,twocolumn,amsmath,amssymb,longbibliography,numbers,floatfix]{revtex4-1}

\usepackage[utf8]{inputenc}
\usepackage{newtxtext}
\usepackage[upint]{newtxmath}
\usepackage{microtype}
\usepackage{textcomp}
\usepackage{eucal}
\usepackage{bm}
\usepackage{siunitx}
\usepackage{comment}

\usepackage{enumerate}
\usepackage{amsfonts}
\usepackage{amsmath}
\usepackage{amssymb}
\usepackage{color}
\usepackage{soul}

\usepackage{graphicx}

\usepackage[colorlinks,allcolors=blue]{hyperref}
\usepackage[capitalize]{cleveref}

\definecolor{DarkRed}{rgb}{0.65,0,0}%
\definecolor{Green}{rgb}{0,0.3,0.3}
\definecolor{Purple}{rgb}{0.3,0,0.65}
\definecolor{Red}{rgb}{1,0,0}
\definecolor{Blue}{rgb}{0,0,0.85}

\newcommand{\Real}{{\mathrm{Re}}}   
\newcommand{\ve}[1]{\boldsymbol{#1}}

\newcommand{\vece}{\ve{e}}

\newcommand{\be}{\begin{equation}}
\newcommand{\ee}{\end{equation}}

\newcommand\combullet{%
  \mathrel{{\ooalign{\hss\raisebox{-0.3ex}{$,$}\hss\cr\raisebox{0.3ex}{$\bullet$}}}}
}

\newcommand{\vehsigma}{\hat{\ve{\sigma}}}

\newcommand{\vep}{\ve{p}}
\newcommand{\veq}{\ve{q}}

\newcommand{\hrhot}{\hat{\rho}_{3}}
\newcommand{\hrhoz}{\hat{\rho}_{0}}
\newcommand{\veR}{\ve{R}}
\newcommand{\ver}{\ve{r}}

\newcommand{\prlsection}[1]{\textit{#1}.\kern0.05em---\kern0.05em\ignorespaces}

\begin{document}
\title{Spin injection and spin relaxation in odd-frequency superconductors}
\author{Lina G. Johnsen}
\email{lina.g.johnsen@ntnu.no}
\affiliation{Center for Quantum Spintronics, Department of Physics, \\ Norwegian University of Science and Technology, NO-7491 Trondheim, Norway}
\author{Jacob Linder}
\affiliation{Center for Quantum Spintronics, Department of Physics, \\ Norwegian University of Science and Technology, NO-7491 Trondheim, Norway}
\date{\today}
\begin{abstract}
The spin transport inside an odd-frequency spin-triplet superconductor differs from that of a conventional superconductor due to its distinct symmetry properties. We study spin transport inside an emergent odd-frequency superconductor by replacing the spin-singlet gap matrix in the Usadel equation with a matrix representing spin-triplet pairing that is odd under inversion of energy. We show that the peculiar nature of the density of states allows for an even larger spin injection than in the normal-state. Moreover, when the odd-frequency pairing inherits its temperature dependence from a conventional superconductor through the proximity effect, the density of states can transition from gapless to gapped as the temperature decreases. At the transition point, the spin accumulation inside the odd-frequency superconductor is peaked and larger than in the normal-state. While the spin-flip scattering time is known to decrease below the superconducting transition temperature in conventional superconductors, we find that the same is true for the spin-orbit scattering time in odd-frequency superconductors. This renormalization is particularly large for energies close to the gap edge, if such a gap is present. 
\end{abstract}
\maketitle


Odd-frequency superconductivity possesses the same robustness against disorder as conventional superconductivity, while allowing for the existence of Cooper pairs that can carry a net spin.  These properties are inherited from the $s$-wave spin-triplet symmetry of the Cooper pairs \cite{anderson_jpcs_59}, and makes odd-frequency superconductors interesting candidates for dissipationless spin transport \cite{bergeret_rmp_05,buzdin_rmp_05,tanaka_jpsj_12,linder_rmp_19}. 
Subsequent to the first proposal of odd-frequency pairing as an allowed symmetry of the superconducting state \cite{berezinskii_jetp_74}, a number of structures have been suggested for realizing odd-frequency superconductivity \cite{eschrig_rpp_15}. Among these are superconductor/ferromagnet hybrids where conventional Cooper pairs are transformed into odd-frequency spin-triplets in the presence of the ferromagnetic exchange field \cite{bergeret_prl_01}. These can penetrate deep into the ferromagnet when \textit{e.g.} 
noncolinear magnetization alignment \cite{volkov_prl_03,bergeret_prb_03,khaire_prl_10,robinson_sc_10,robinson_prl_10,klose_prl_12}, 
inhomogenous magnetization \cite{eschrig_prl_03,keizer_n_06,anwar_apl_12,banerjee_nc_14}, or
spin-orbit coupling \cite{bergeret_prl_13,bergeret_prb_14,costa_arxiv_21} is used to form equal-spin triplet pairs unaffected by the Zeeman spin-splitting.
By now, signatures of odd-frequency triplets have been observed in many different structures, \textit{e.g.} through modulation of the 
superconducting critical temperature \cite{leksin_prl_12,wang_prb_14,banerjee_prb_18},
density of states (DOS) \cite{dibernardo_nc_15,diesch_nc_18}, and
magnetic anisotropy \cite{gonzalez-ruano_prb_20,gonzalez-ruano_arxiv_21}, and through
observation of long-range supercurrents in Josephson junctions \cite{keizer_n_06,khaire_prl_10,robinson_sc_10,robinson_prl_10,klose_prl_12,anwar_apl_12,banerjee_nc_14}, 
and the paramagnetic Meissner effect \cite{yokoyama_prl_11,dibernardo_prx_15}.

\begin{figure}[b]
    \centering
    \includegraphics[width=\columnwidth]{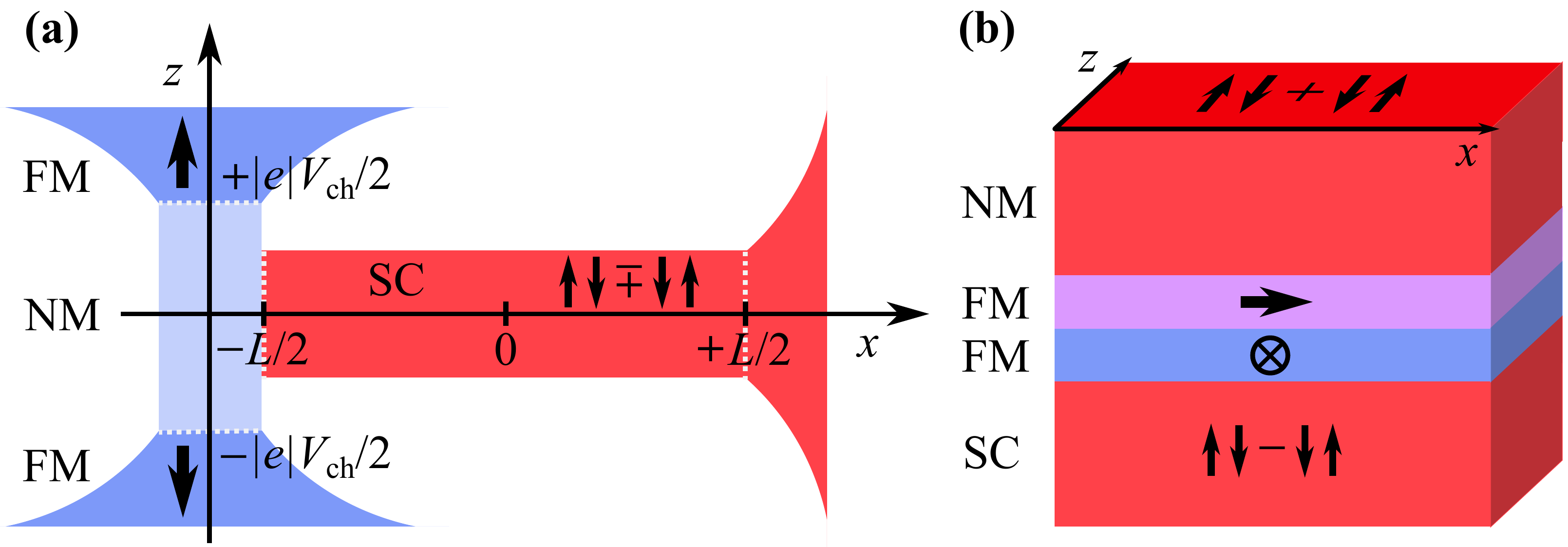}
    \caption{(a) We study the non-equilibrium spin accumulation in a conventional spin-singlet $(\uparrow\downarrow-\downarrow\uparrow)_z$ and odd-frequency spin-triplet $(\uparrow\downarrow+\downarrow\uparrow)_z$ superconductor (SC) upon applying a spin-dependent voltage $|e|V$ to an adjacent normal-metal (NM). The spin-dependent voltage has opposite sign for spin-up and spin-down electrons, and can be induced from an electric voltage $|e|V_{\text{ch}}$ applied between two oppositely oriented ferromagnets (FM). The applied electric voltage $|e|V_{\text{ch}}$ is in general not equal to the induced spin-dependent voltage $|e|V$ in the NM contact. The FMs are polarized along the $z$ axis so that the spins injected into the SC cannot be carried by the Cooper pairs.
    The injected spins are relaxed by spin-flip and spin-orbit scattering until equilibrium is reached a distance $L$ from the NM contact.
    (b)~We suggest inducing odd-frequency superconductivity through proximity to a conventional superconductor. Spin-singlet Cooper pairs are partially converted into triplets as they leak from a conventional superconductor into a ferromagnet. Upon leaking through a second sufficiently thick ferromagnet magnetized perpendicularly to the first one, only spin-triplets survive \cite{leksin_prl_12,jara_prb_14,wang_prb_14}. In the highly disordered materials considered here, only $s$-wave pairing can be present \cite{anderson_jpcs_59,bergeret_prl_01}. The remaining triplet pairing then transforms the adjacent normal-metal into an emergent odd-frequency superconductor.}
    \label{fig:model}
\end{figure}

From a symmetry point of view, the odd-frequency superconducting pairing differs from the conventional one by its spin-triplet symmetry leaving it invariant under exchange of spin coordinates, and an odd parity with respect to exchange of time coordinates for the electrons in the Cooper pair.
While the $s$-wave symmetry ensures robustness under regular impurity scattering for both conventional and odd-frequency superconductors, the former is expected to be less robust to magnetic impurities and the latter to spin-orbit scattering \cite{bergeret_rmp_05,demler_prb_97,rachataruangsit_pc_07}.
As was first discussed in the context of the proximity effect in superconductor/ferromagnet structures \cite{buzdin_prb_00}, another characteristic of odd-frequency pairing is that it alters the local DOS. In an odd-frequency superconductor, the DOS can follow an energy dependence similar to that of the conventional superconductor with a gap around the Fermi energy. However, another possibility is that the DOS is gapless and peaked at zero energy \cite{tanaka_prl_07,sukhachov_prb_19}.
These properties are essential for describing the spin transport inside the odd-frequency superconductor.

In conventional superconductors, Cooper pairs are spin-less and quasi-particles are responsible for the spin transport \cite{johnson_apl_94,zhao_prb_95}. Therefore, spin injection is blocked at energies below the gap edge.
The onset of superconductivity also causes the spin-flip scattering length to become energy dependent. For energies close to the gap edge, there is a giant renormalization of the spin-flip scattering length causing a rapid decrease in the spin accumulation inside the superconductor \cite{morten_prb_04}.
Additionally, the magnetic impurities causes a weakening of the superconducting gap \cite{morten_prb_05}.
The spin-orbit scattering time is not renormalized by conventional superconductivity and remains equal to its normal state value \cite{morten_prb_04}.

In this work, we instead consider the non-equilibrium spin accumulation in an emergent odd-frequency superconductor in the presence of spin-flip and spin-orbit scattering. We compare our results to the conventional case.
Our approach is to consider the Usadel equation for a conventional superconductor, and then to replace the conventional gap matrix with a contribution with a spin-triplet symmetry and odd parity under inversion of energy.
We study the system shown in Fig.~\ref{fig:model}(a), where spin is injected into the odd-frequency or conventional superconductor from a normal-metal contact under an applied spin voltage. The polarization axis of the injected spin is chosen so that the spin transport is carried by quasi-particles only.
Although odd-frequency superconductivity has not been found to exist intrinsically in materials, it can be induced by the proximity effect. One way of doing this is presented in Fig.~\ref{fig:model}(b), where leakage of Cooper pairs through two misaligned ferromagnets effectively converts a normal-metal into an emergent odd-frequency spin-triplet superconductor. Therefore, our predictions can be tested experimentally in a hybrid structure.

The odd-frequency superconductivity does not renormalize the spin-flip scattering time caused by magnetic impurities compared to the normal state \cite{bergeret_rmp_05,demler_prb_97,rachataruangsit_pc_07}. Instead, the spin-orbit scattering length decreases below the superconducting critical temperature. We find that when a gap is present in the DOS, there is a giant renormalization of the spin-orbit scattering length at the gap edge, similar to the renormalization of the spin-flip scattering length in conventional superconductors.
Moreover, we find that the distinct features of the DOS \cite{sukhachov_prb_19} causes the temperature dependence of the non-equilibrium spin accumulation to behave qualitatively different from what is expected for a conventional superconductor. It presents a peak when the DOS transitions from peaked to gapped as the temperature decreases. The possibility of a high DOS at low energies also opens the possibility of a higher spin injection than in the normal state. 

\section{Theoretical framework}

\subsection{Quasi-classical theory for conventional diffusive superconductors}

Our approach will be to generalize the the quasiclassical theory for a diffusive conventional superconductor \cite{eckern_jltp_81,belzig_sm_99} in order to describe odd-frequency spin-triplet pairing \cite{tanaka_prl_07,sukhachov_prb_19}.
The motivation behind using this approach is that writing down a microscopic Hamiltonian for the odd-frequency pairing would require adding a time dependence to the electron creation and annihilation operators, which greatly adds to the complexity of the problem. However, we know that the odd-frequency pairing has an even parity under spin inversion and an odd parity under inversion of energy. We can then generalize the result for the conventional pairing so that the pairing satisfies the desired symmetry relations. This comes at the cost of not knowing the gap equation for the odd-frequency pairing.

The impurity-averaged quasi-classical Green's function $\check{g}_{\text{av}}^{\text{s}}(\veR,\epsilon)$ of a diffusive conventional superconductor can be described by the Usadel equation
\be
\begin{split}
    &\nabla_{\veR}\cdot\check{\ve{I}}(\veR,\epsilon)=i[\check{\sigma}(\veR,\epsilon),\check{g}_{\text{av}}^{\text{s}}(\veR,\epsilon)].
\end{split}
\label{eq:usadel}
\ee
Its underlying assumptions and derivation starting from a continuum model is described in Appendix~\ref{appendix:Usadel}.
The Green's function is defined in Keldysh space and has the matrix structure
\begin{align}
    \check{g}_{\text{av}}^{\text{s}}(\veR,\epsilon)=
    \begin{pmatrix}
       [\hat{g}_{\text{av}}^{\text{s}}(\veR,\epsilon)]^{\text{R}} & [\hat{g}_{\text{av}}^{\text{s}}(\veR,\epsilon)]^{\text{K}}\\
        0 & [\hat{g}_{\text{av}}^{\text{s}}(\veR,\epsilon)]^{\text{A}}
    \end{pmatrix},
    \label{eq:greens_function_keldysh_space}
\end{align}
where $[\hat{g}_{\text{av}}^{\text{s}}(\veR,\epsilon)]^{\text{R}}$, $[\hat{g}_{\text{av}}^{\text{s}}(\veR,\epsilon)]^{\text{A}}$, and $[\hat{g}_{\text{av}}^{\text{s}}(\veR,\epsilon)]^{\text{K}}$ are the impurity-averaged quasi-classical retarded, advanced and Keldysh Green's functions, respectively. The check denotes $8\times8$ matrices in Keldysh space, where $\check{\rho}_0$ is the unit matrix, while the hat denotes $4\times4$ matrices in Nambu~$\otimes$~spin space. 
We have defined a matrix current
\be
\check{\ve{I}}(\veR,\epsilon)=-D\check{g}_{\text{av}}^{\text{s}}(\veR,\epsilon)\nabla_{\veR}\check{g}_{\text{av}}^{\text{s}}(\veR,\epsilon),
\ee
where $D=\tau v_F^2 /3$ is the diffusion coefficient. The diffusion coefficient is determined by the Fermi velocity $v_{\text{F}}=p_F /m$, and the scattering time associated with scattering on non-magnetic impurities $\tau = \big[2\pi n N_0 \big<|u(\ve{e}_{p_F}-\ve{e}_{q_F})|^2 \big>_{p_F ,q_F}\big]^{-1}$. Here, $n$ is the density of non-magnetic impurities, $N_0$ is the DOS at the Fermi level, and  $u(\ve{e}_{p_F}-\ve{e}_{q_F})$ is the scattering potential of a single non-magnetic impurity. The scattering potential is averaged over the all possible directions of the momenta $\vep_{\text{F}}=p_F \vece_{p_F}$ and $\veq_{\text{F}}=q_F \vece_{q_F}$, where $\vece_{q_F}$ and $\vece_{p_F}$ are unit vectors. The self-energy matrix
$\check{\sigma}(\veR,\epsilon)=\hat{\sigma}_0 (\epsilon)  +\check{\sigma}_{\text{sf}}(\veR,\epsilon)+\check{\sigma}_{\text{so}}(\veR,\epsilon)+\hat{\sigma}_{\text{sc}}^{\text{S}} (\veR)$ contains the contributions
\begin{align}
    \hat{\sigma}_0 (\epsilon)&=\epsilon\hrhot,\\
    \check{\sigma}_{\text{sf}}(\veR,\epsilon)&=(i/8\tau_{\text{sf}})\vehsigma\cdot \check{g}_{\text{av}}^{\text{s}}(\veR,\epsilon)\vehsigma,\\
    \check{\sigma}_{\text{so}}(\veR,\epsilon)&=(i/8\tau_{\text{so}})\hrhot\vehsigma\cdot \check{g}_{\text{av}}^{\text{s}}(\veR,\epsilon)\hrhot\vehsigma,\\
    \hat{\sigma}_{\text{sc}}^{\text{S}}(\veR)&=\hat{\Delta}^{\text{S}}(\veR).
\end{align}
Above, $\epsilon$ is the quasi-particle energy, $\veR$ is the center-of-mass coordinate associated with the Green's function, $\hrhot=\text{diag}(1,1,-1,-1)$, and $\hat{\ve{\sigma}}=\text{diag}(\ve{\sigma},\ve{\sigma}^* )$, where $\ve{\sigma}$ is the vector of Pauli matrices. 
The self-energies $\check{\sigma}_{\text{sf}}(\veR,\epsilon)$ and $\check{\sigma}_{\text{so}}(\veR,\epsilon)$ describe the spin-flip scattering on magnetic impurities and the spin-orbit scattering on non-magnetic impurities, respectively. The respective scattering times are given by $\tau_{\text{sf}}=\big[8\pi n_{\text{m}} N_0 \big<|u_{\text{m}} (\ve{e}_{p_F}-\ve{e}_{q_F})|^2 \big>_{p_F ,q_F }S(S+1)/3\big]^{-1}$ and $\tau_{\text{so}}=9\tau/(8\alpha^2 p_F^4 )$. 
Here, $n_{\text{m}}$ is the density of the magnetic impurities, $u_{\text{m}}(\ve{e}_{p_F}-\ve{e}_{q_F})$ and $S$ are the scattering potential and the spin of a single magnetic impurity, and $\alpha$ is the Rashba parameter.
We have assumed that $\tau_{\text{so}},\tau_{\text{sf}}\gg\tau$ so that scattering on non-magnetic impurities dominates over the spin-orbit and spin-flip scattering.
The spin-singlet superconducting pairing is described by the gap matrix
\begin{align}
\hat{\Delta}^{\text{S}}=\text{antidiag}\{\Delta^{\text{S}},-\Delta^{\text{S}},
[\Delta^{\text{S}}]^* ,-[\Delta^{\text{S}}]^* \}.
\end{align}
In this work, we will assume the spin-singlet superconducting gap $\Delta^{\text{S}}$ to be spatially independent and follow a standard Bardeen-Cooper-Schrieffer temperature dependence given by $\Delta^{\text{S}}(T)=\Delta_0 f(T)$, where
\begin{align}
    \Delta_0&=1.76T_c ,\\
   f(T) &=\tanh\bigg(1.74\sqrt{\frac{T_c}{T}-1}\:\bigg),
\end{align}
are the zero-temperature gap and the temperature dependence of the gap, respectively. Above, $T_c$ is the superconducting critical temperature. The assumption that the superconducting gap, and in particular its phase, is spatially independent holds as long as we only consider spin transport. This is because spin can only be carried by the quasi-particles. Charge can on the other hand be carried by both quasi-particles and Cooper pairs. This causes a conversion between quasi-particle and Cooper pair transport that leads to a renormalization of the gap and makes a spatially dependent phase of the order parameter necessary.

\subsection{Model for odd-frequency superconductivity}

In order to describe an odd-frequency spin-triplet superconductor, we replace the spin-singlet contribution to the Usadel equation $\hat{\sigma}^{\text{S}}$ by an energy-dependent contribution $\hat{\sigma}^{\text{T}}(\epsilon)=\hat{\Delta}^{\text{T}}(\epsilon)$ which has an odd parity with respect to inversion of energy. Providing the order parameter with a dependence on $\epsilon$ in this way produces the correct relation between the retarded and advanced Green function required for odd-frequency pairing, $f^R_{\alpha\beta}(\epsilon) = - f^A_{\alpha\beta}(-\epsilon)$, corresponding to an odd parity with respect to exchange of time coordinates \cite{linder_rmp_19}. We also alter the structure of the gap matrix
\begin{align}
\hat{\Delta}^{\text{T}}(\epsilon)=\text{antidiag}&\{\Delta^{\text{T}}(\epsilon),\Delta^{\text{T}}(\epsilon),
-[\Delta^{\text{T}} (\epsilon)]^* ,-[\Delta^{\text{T}} (\epsilon)]^* \}
\end{align}
in order to describe spin-triplet pairing. 
We model the spin-triplet pairing by two different plausible models \cite{sukhachov_prb_19}
\begin{align}
    \Delta^{\text{T}}(\epsilon,T)&=\frac{ C f(T) \epsilon}{1+\left(\frac{C\epsilon}{2\Delta_{\text{max}}}\right)^2},\label{eq:type1}\\
    \Delta^{\text{T}}(\epsilon,T)&=\frac{Cf(T)  \epsilon}{\sqrt{1+\left(\frac{C\epsilon}{\Delta_{\infty}}\right)^2}}\label{eq:type2},
\end{align}
giving rise to similar results.
Equation~\eqref{eq:type1} describes a pairing that has a linear form $Cf(T) \epsilon$ for small energies, reaches it maximum $\Delta_{\text{max}}$, and then decays as $\sim1/\epsilon$ for large energies. Equation~\eqref{eq:type2} describes a pairing that has the same linear form for small energies, and that approaches a constant value $\Delta_{\infty}$ for large energies. We set the maximum pairing $\Delta_{\text{max}}$ and $\Delta_{\infty}$ of the above models equal to the zero-temperature singlet gap $\Delta_0$. As can be seen from the above equations, we have assumed the temperature dependence to be the same as for the singlet pairing. This is because when the odd-frequency triplet paring is produced by the proximity effect as described in Fig.~\ref{fig:model}(b), the temperature dependence is inherited from the original singlet pairing.
Both of the above models produce a gapped DOS similar to that of a spin-singlet superconductor for $Cf(T) >1$. The gap is of magnitude $(2\Delta_{\text{max}}/C)\sqrt{Cf(T) - 1}$ for the pairing in Eq.~\eqref{eq:type1} and $(\Delta_{\infty}/C)\sqrt{[Cf(T)]^2 -1}$ for the pairing in Eq.~\ref{eq:type2}. For $0<Cf(T) \leq1$, the DOS is instead gapless and peaked around $\epsilon=0$ \cite{tanaka_prl_07,sukhachov_prb_19}.
Note that the assumption that the pairing is spatially independent also holds for the odd-frequency pairing considered here. The triplet pairs considered ($S_z=0$) cannot carry any spin supercurrent polarized in the $z$ direction, and thus we may consider a spatially homogeneous order parameter.

\subsection{The kinetic equations and the non-equilibrium spin accumulation}

The Usadel equation is subject to a normalization condition
\be
\check{g}_{\text{av}}^{\text{s}}(\veR,\epsilon)\check{g}_{\text{av}}^{\text{s}}(\veR,\epsilon)=\check{\rho}_0 
\ee
for the quasi-classical Green's function.
It follows from the normalization condition that the quasi-classical Keldysh Green's function can be written it terms of the retarded and advanced Green's functions as
\begin{align}
[\hat{g}_{\text{av}}^{\text{s}} (\veR,\epsilon)]^K = &[\hat{g}_{\text{av}}^{\text{s}} (\veR,\epsilon)]^R  \hat{h}(\veR,\epsilon)\notag\\
&-\hat{h}(\veR,\epsilon)  [\hat{g}_{\text{av}}^{\text{s}} (\veR,\epsilon)]^A ,
\end{align}
where $\hat{h}(\veR,\epsilon)$ is the distribution matrix. Moreover, it follows from the definitions of the retarded and advanced Green's functions that these are related by
$[\hat{g}_{\text{av}}^{\text{s}} (\veR,\epsilon)]^A =-\{\hrhot [\hat{g}_{\text{av}}^{\text{s}} (\veR,\epsilon)]^R \hrhot  \}^{\dagger}$. In order to solve the Usadel equation for our system, we therefore only need expressions for the distribution matrix and the retarded Green's function. 
We assume that the distribution matrix is diagonal, and write it as
\begin{align}
\hat{h}(\veR,\epsilon)=&\phantom{+}\hrhoz h_{\text{L}}(\veR,\epsilon) +\hrhot h_{\text{T}}(\veR,\epsilon)\notag\\
&+\sum_i (\vehsigma)_i h_{\text{LS}}^i (\veR,\epsilon)+\sum_i \hrhot(\vehsigma)_i h_{\text{TS}}^i (\veR,\epsilon),
\end{align}
where $i\in\{x,y,z\}$ refers to the spin projection axis.
Above, $h_{\text{L}}(\veR,\epsilon)$, $h_{\text{T}}(\veR,\epsilon)$, $h_{\text{LS}}^i (\veR,\epsilon)$, and $h_{\text{TS}}^i (\veR,\epsilon)$ are the energy, charge, spin-energy, and spin distribution functions, respectively.
We define corresponding current densities
\begin{align}
    \ve{j}_{\text{L}} (\veR,\epsilon)&= \text{Tr} \{\hat{\ve{I}}^K (\veR,\epsilon)\}/4,\\
    \ve{j}_{\text{T}} (\veR,\epsilon)&= \text{Tr} \{\hrhot\hat{\ve{I}}^K (\veR,\epsilon)\}/4,\\
    \ve{j}_{\text{LS}}^i (\veR,\epsilon)&= \text{Tr} \{(\vehsigma)_i \hat{\ve{I}}^K (\veR,\epsilon)\}/4,\\
    \ve{j}_{\text{TS}}^i (\veR,\epsilon)&= \text{Tr} \{\hrhot(\vehsigma)_i \hat{\ve{I}}^K (\veR,\epsilon)\}/4
\end{align}
in terms of the Keldysh part of the current matrix.
We set the retarded Green's function for the spin-singlet (spin-triplet) superconductor equal to its equilibrium solution,
\begin{align}
\{[\hat{g}_{\text{av}}^{\text{s}} (\epsilon)]^R\}^{\text{S}(\text{T})} =&[\hat{\rho}_3 \epsilon +\hat{\Delta}^{\text{S}(\text{T})}(\epsilon)]I^{\text{S}(\text{T})}(\epsilon),\\
I^{\text{S}(\text{T})}(\epsilon)=&\frac{\text{sgn}(\epsilon)\Theta\left(\epsilon^2 -|\Delta^{\text{S}(\text{T})}(\epsilon)|^2 \right)}{\sqrt{\epsilon^2 -|\Delta^{\text{S}(\text{T})}(\epsilon)|^2 }}\notag\\ 
&-\frac{i\Theta\left(|\Delta^{\text{S}(\text{T})}(\epsilon)|^2 -\epsilon^2 \right)}{\sqrt{|\Delta^{\text{S}(\text{T})}(\epsilon)|^2 -\epsilon^2 }},
\end{align}
throughout the superconducting region. Above, $\Theta(\epsilon)$ is the Heaviside step function. 
We have neglected the influence of magnetic and spin-orbit impurity scattering on the retarded Green's function and instead study how the impurity scattering affects the spin distribution function. 
A study of how the abovementioned scattering changes the Green's function would require a self-consistent solution for the superconducting pairing and would reveal a renormalization of the superconducting gap. A self-consistent solution is not possible for the spin-triplet superconductor for which the gap equation is unknown. Although a self-consistent solution would not reveal a mixing between conventional and odd-frequency paring in the present framework,  it have been shown to occur close to single magnetic impurities in clean superconductors \cite{kuzmanovski_prb_20,perrin_prl_20}. If such a mixing were present, there would be a contribution from both types of pairing to the non-equilibrium spin-accumulation.

Focusing now on the spin transport, we insert the equilibrium retarded Green's function and the definition of the distribution functions into the Keldysh component of the Usadel equation. For the spin-singlet (spin-triplet) superconducting pairing, we find a relation 
\begin{align}
    &\nabla_{\veR}\cdot[\ve{j}_{\text{TS}}^z (\veR,\epsilon)]^{\text{S}(\text{T})}
    =-2\alpha_{\text{TSTS}}^{\text{S}(\text{T})}(\epsilon)[h_{\text{TS}}^z (\veR,\epsilon)]^{\text{S}(\text{T})} ,\label{eq:nabla_dot_jTS}\\
    &\alpha_{\text{TSTS}}^{\text{S}}(\epsilon)=
    \left(\frac{1}{\tau_{\text{so}}}+\frac{1}{\tau_{\text{sf}}}\frac{\epsilon^2 +|\Delta|^2 }{\epsilon^2 -|\Delta|^2}\right)\Theta\big(\epsilon^2 -|\Delta|^2 \big),\\
    &\alpha_{\text{TSTS}}^{\text{T}}(\epsilon)=
    \left(\frac{1}{\tau_{\text{sf}}}+\frac{1}{\tau_{\text{so}}}\frac{\epsilon^2 +|\Delta(\epsilon)|^2 }{\epsilon^2 -|\Delta(\epsilon)|^2}\right)\Theta\big(\epsilon^2 - |\Delta(\epsilon)|^2\big),
\end{align}
between the spin current density $j_{\text{TS}}^{z}(\veR,\epsilon)$ and the spin distribution function $h_{\text{TS}}^{z}(\veR,\epsilon)$. 
Notice that while spin-singlet superconductivity renormalizes the spin-flip scattering time, the odd-frequency spin-triplet superconductivity instead renormalizes the spin-orbit scattering time. For a gapped triplet superconductor ($Cf(T) >1$), we see from the above expression that there occurs a giant renormalization at the gap edge $\epsilon \to \Delta(\epsilon)$ causing rapid spin-orbit relaxation.
From the definition of the spin current density, we find that
\begin{align}
    [\ve{j}_{\text{TS}}^z (\veR,\epsilon)]^{\text{S}(\text{T})} &=-2D_{\text{L}}^{\text{S}(\text{T})}(\epsilon)\nabla_{\veR}[h_{\text{TS}}^z (\veR,\epsilon)]^{\text{S}(\text{T})},\label{eq:jTS_expression}\\
    D_{\text{L}}^{\text{S}}(\epsilon)&=
    D\Theta\big(\epsilon^2 -|\Delta|^2 \big),\\
    D_{\text{L}}^{\text{T}}(\epsilon)&=
    D\Theta\big(\epsilon^2 -|\Delta(\epsilon)|^2 \big).
\end{align}
In order to study the spin distribution $[h_{\text{TS}}^z (\veR,\epsilon)]^{\text{S}(\text{T})}$ inside a singlet (triplet) superconductor under spin injection, we introduce for simplicity transparent boundaries to a normal-metal with a spin voltage $V_{\uparrow}=-V_{\downarrow}= V/2$ at position $x=-L/2$. Using more realistic tunneling boundary conditions simply diminishes the magnitude of the spin injection, regardless of whether we consider a conventional superconductor or an odd-frequency superconductor, and does not change any of our conclusions. We assume the spin injected into the singlet or triplet superconductor from the normal-metal to have relaxed completely at $x=L/2$. This corresponds to the system introduced in Fig.~\ref{fig:model}(a). This situation can be described by the boundary conditions
\begin{align}
    h_{\text{TS}}^z (-L/2,\epsilon) =&\frac{1}{2}\left[\tanh\left(\frac{\epsilon+eV_{\uparrow}}{2T}\right)-\tanh\left(\frac{\epsilon+eV_{\downarrow}}{2T}\right)\right],\label{eq:h_at_-L/2}\\
    h_{\text{TS}}^z (L/2,\epsilon) =&0,
\end{align}
where the temperature $T$ is constant throughout the material.
Solving Eqs.~\eqref{eq:nabla_dot_jTS} and~\eqref{eq:jTS_expression} with these boundary conditions, we find that the spin distribution function for the singlet (triplet) superconductor is given by
\begin{widetext}
\begin{align}
    [h_{\text{TS}}^z (x,\epsilon)]^{\text{S}(\text{T})}&=
    \frac{1}{2}h_{\text{TS}}^{z}(-L/2,\epsilon)[H_{\text{TS}}^{z}(x,\epsilon)]^{\text{S}(\text{T})}\Theta\big(\epsilon^2 -|\Delta^{\text{S}(\text{T})}(\epsilon)|^2 \big),\label{eq:h}\\
    [H_{\text{TS}}^{z}(x,\epsilon)]^{\text{S}}&=\left\{\frac{\cosh\left(\sqrt{\frac{1}{l_{\text{so}}^2}+\frac{1}{l_{\text{sf}}^2}\frac{\epsilon^2 +|\Delta^{\text{S}}|^2}{\epsilon^2 -|\Delta^{\text{S}}|^2}}x\right)}{\cosh\left(\sqrt{\frac{1}{l_{\text{so}}^2}+\frac{1}{l_{\text{sf}}^2}\frac{\epsilon^2 +|\Delta^{\text{S}}|^2}{\epsilon^2 -|\Delta^{\text{S}}|^2}}\frac{L}{2}\right)}
    -\frac{\sinh\left(\sqrt{\frac{1}{l_{\text{so}}^2}+\frac{1}{l_{\text{sf}}^2}\frac{\epsilon^2 +|\Delta^{\text{S}}|^2}{\epsilon^2 -|\Delta^{\text{S}}|^2}}x\right)}{\sinh\left(\sqrt{\frac{1}{l_{\text{so}}^2}+\frac{1}{l_{\text{sf}}^2}\frac{\epsilon^2 +|\Delta^{\text{S}}|^2}{\epsilon^2 -|\Delta^{\text{S}}|^2}}\frac{L}{2}\right)}\right\},\label{eq:HS}\\
    [H_{\text{TS}}^{z}(x,\epsilon)]^{\text{T}}&=\left\{\frac{\cosh\left(\sqrt{\frac{1}{l_{\text{sf}}^2}+\frac{1}{l_{\text{so}}^2}\frac{\epsilon^2 +|\Delta^{\text{T}}(\epsilon)|^2}{\epsilon^2 -|\Delta^{\text{T}}(\epsilon)|^2}}x\right)}{\cosh\left(\sqrt{\frac{1}{l_{\text{sf}}^2}+\frac{1}{l_{\text{so}}^2}\frac{\epsilon^2 +|\Delta^{\text{T}}(\epsilon)|^2}{\epsilon^2 -|\Delta^{\text{T}}(\epsilon)|^2}}\frac{L}{2}\right)}
    -\frac{\sinh\left(\sqrt{\frac{1}{l_{\text{sf}}^2}+\frac{1}{l_{\text{so}}^2}\frac{\epsilon^2 +|\Delta^{\text{T}}(\epsilon)|^2}{\epsilon^2 -|\Delta^{\text{T}}(\epsilon)|^2}}x\right)}{\sinh\left(\sqrt{\frac{1}{l_{\text{sf}}^2}+\frac{1}{l_{\text{so}}^2}\frac{\epsilon^2 +|\Delta^{\text{T}}(\epsilon)|^2}{\epsilon^2 -|\Delta^{\text{T}}(\epsilon)|^2}}\frac{L}{2}\right)}\right\}.\label{eq:HT}
\end{align}
\end{widetext}
We have defined the normal-state spin-flip and spin-orbit relaxation lengths $l_{\text{sf}}=\sqrt{D\tau_{\text{sf}}}$ and $l_{\text{so}}=\sqrt{D\tau_{\text{so}}}$.
 The non-equilibrium spin-accumulation 
\be
[\mu^{z} (x)]^{\text{S}(\text{T})} = -\frac{1}{N_0}\int_{-\infty}^{\infty} d\epsilon\:N^{\text{S}(\text{T})}(\epsilon)[h_{\text{TS}}^z (x,\epsilon)]^{\text{S}(\text{T})}
\ee
is determined by the spin distribution function given above and the DOS
$N^{\text{S}(\text{T})}(\epsilon)=N_0 \Real(\{[g_{\text{av}}^{\text{s}}(\veR,\epsilon)]^{\text{R}}\}^{\text{S}(\text{T})})$, where $\{[g_{\text{av}}^{\text{s}}(\veR,\epsilon)]^{\text{R}}\}^{\text{S}(\text{T})}=\epsilon I^{\text{S}(\text{T})}(\epsilon)$. 

\section{The non-equilibrium spin accumulation}

\subsection{The density of states}
\label{s:DOS}

\begin{figure}[htb]
    \centering
    \includegraphics[width=\columnwidth]{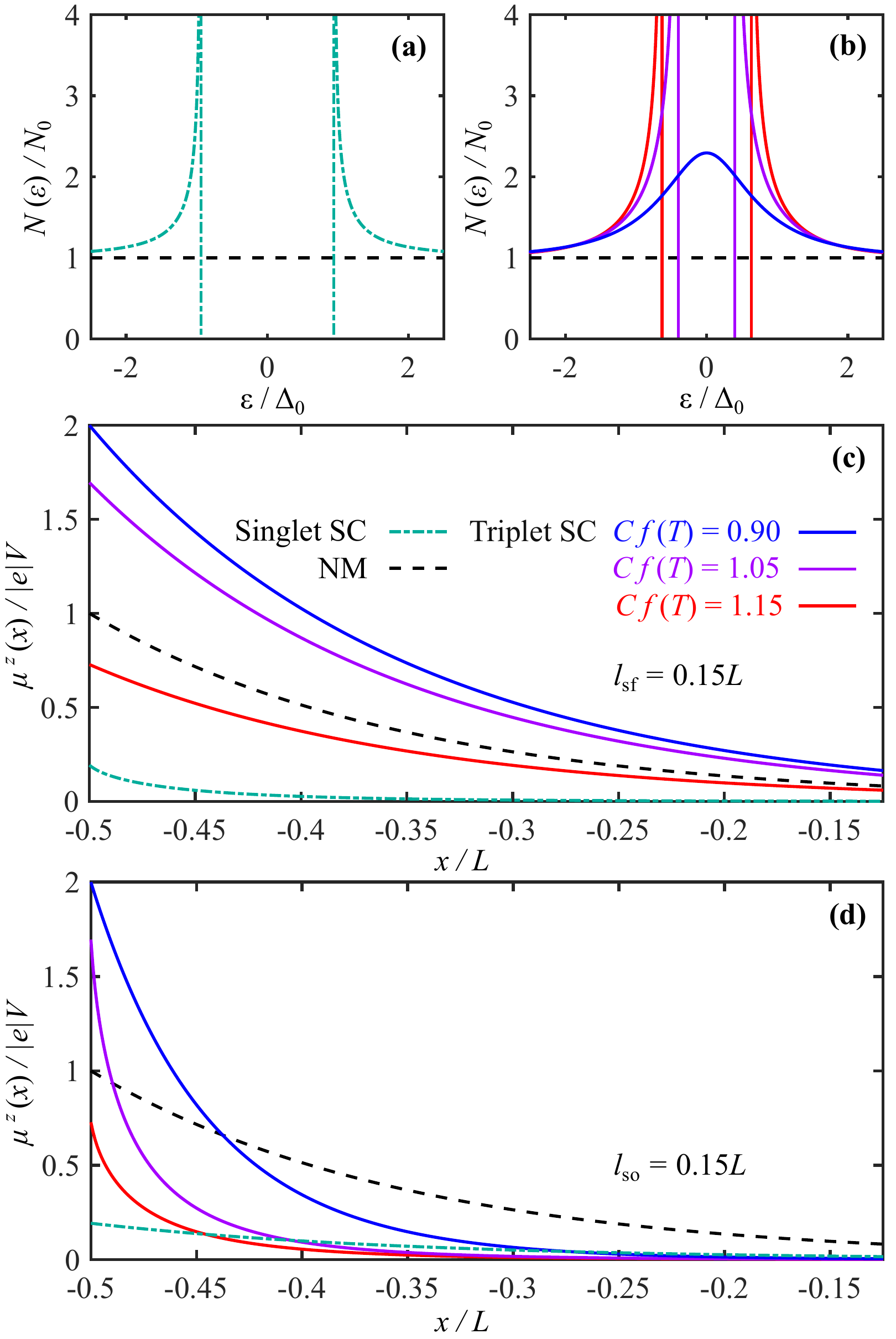}
    \caption{While the DOS for a singlet superconductor (a) is gapped, the DOS of an odd-frequency superconductor (b) can either be peaked at zero temperature (blue) or gapped (purple and red). The non-equilibrium spin-accumulation corresponding to the DOS in panel (b) is shown for spin-flip scattering with $l_{\text{sf}}=0.15L$ (c) and for spin-orbit scattering with  $l_{\text{so}}=0.15L$ (d). All plots correspond to a spin voltage of $|e|V=0.5\Delta_0$ and temperature $T=0.5T_c$. The above corresponds to the pairing type described in Eq.~\eqref{eq:type1}. The second pairing type described in Eq.~\eqref{eq:type2} gives similar results.}
    \label{fig:DOS_and_position}
\end{figure}

We first discuss how the density of states affects the non-equilibrium spin accumulation inside the superconductor. As shown in Fig.~\ref{fig:DOS_and_position}(a) and~(b), the DOS of an odd-frequency superconductor can either be gapped as in the conventional superconductor, or it can be gapless and peaked at zero energy.
When the DOS is gapped, the situation is similar to that of a conventional superconductor. There are no available states below the gap edge, and spin is blocked from entering the superconductor. At the gap edge, the DOS is large thus allowing for a large spin-injection. 
The conventional superconductor always allows for a spin injection that is less than or equal to the spin injection into a normal-metal. This is because the total number of states is conserved. For energies up to a given spin voltage just above the gap edge, there will be fewer available quasi-particle states as the available states have been pushed out of the gap region towards higher energies. 
For the gapped odd-frequency superconductor, the gap is smaller and the peaks at the gap edge broader than in the conventional superconductor. Although this causes the conservation of the total number of states to be broken in our simple model, it has been shown numerically that this problem can be resolved by flanking the peak at the gap edge by a local minimum \cite{sukhachov_prb_19}. Spin voltages that do not allow for spin injection at such high energies can then give rise to a spin injection that is higher than than the spin injection in the normal state.
For a gapless odd-frequency superconductor, the spin injection can be even larger, since the DOS is always larger than in the normal state, except for at the minima appearing at higher energies \cite{sukhachov_prb_19}. 

Since the spin injection into an odd-frequency superconductor can be larger than than in the normal state, the non-equilibrium spin accumulation close to the normal-metal contact can also be larger. This is demonstrated in Fig.~\ref{fig:DOS_and_position}(c) and~(d) for positions close to $x=-0.5L$.
Note that although the additional minima in the DOS are not included in our analytical model, the error is negligible as long as the temperature and spin voltage is sufficiently low. This is because the distribution function of the normal-metal contact (Eq.~\eqref{eq:h_at_-L/2}) becomes negligibly small at the high energies where the minimum appears. For the temperature and spin voltage used in Fig.~\ref{fig:DOS_and_position} the spin distribution function of the normal-metal is a ten (hundred) times smaller than its maximum value at $\epsilon=1.8\Delta_0$ ($\epsilon=1.1\Delta_0 $).

Another important observation is that since the temperature dependence of the triplet pairing is inherited from the original singlet condensate via the proximity effect (Fig.~\ref{fig:model}(b)), the coefficient $Cf(T)$ determining whether the DOS is gapped or gapless is also temperature dependent. The coefficient $f(T)$ is equal to one at zero temperature and zero at the superconducting critical temperature. This means that if the DOS start out as gapped at $T=0$ ($Cf(0) >1 )$ it must transition to a peaked DOS as $Cf(T)$ drops below one for higher temperatures. Moreover, the DOS diverges as $Cf(T)$ approaches one. This results in a spin injection that is larger than in the normal state due to the high number of available states at zero energy, as we will demonstrate below.

\subsection{Spin-flip and spin-orbit impurity scattering}

\begin{figure}[tb]
    \centering
    \includegraphics[width=\columnwidth]{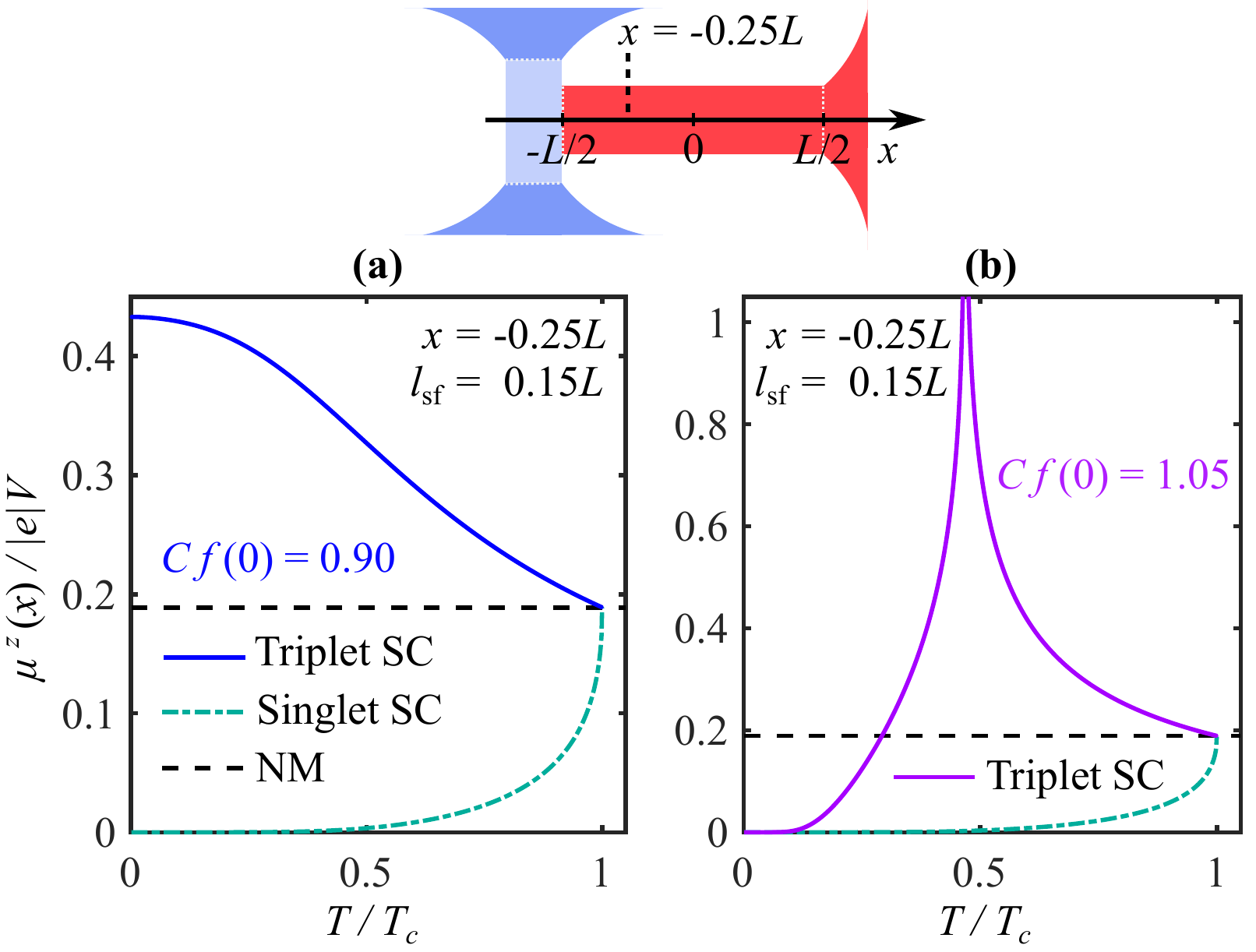}
    \caption{The non-equilibrium spin accumulation is plotted as a function of temperature in the presence of spin-flip scattering for $l_{\text{sf}}=0.15L$. Panel (a) correspond to a gapless DOS where $0\leq Cf(T) \leq0.90$ for all temperatures. Panel (b) correspond to a DOS that is gapped at low temperatures ($1.05>Cf(T) >1$), and gapless at higher temperatures ($0\leq Cf(T)\leq1$). Both are measured at a distance $0.25L$ away from the normal-metal contact and correspond to an applied spin voltage of $|e|V=0.1\Delta_0$. The triplet pairing follows the model described by Eq.~\ref{eq:type1}, however the model described by Eq.~\ref{eq:type2} gives similar results.}
    \label{fig:temperature_sf}
\end{figure}
\begin{figure}[htb]
    \centering
    \includegraphics[width=\columnwidth]{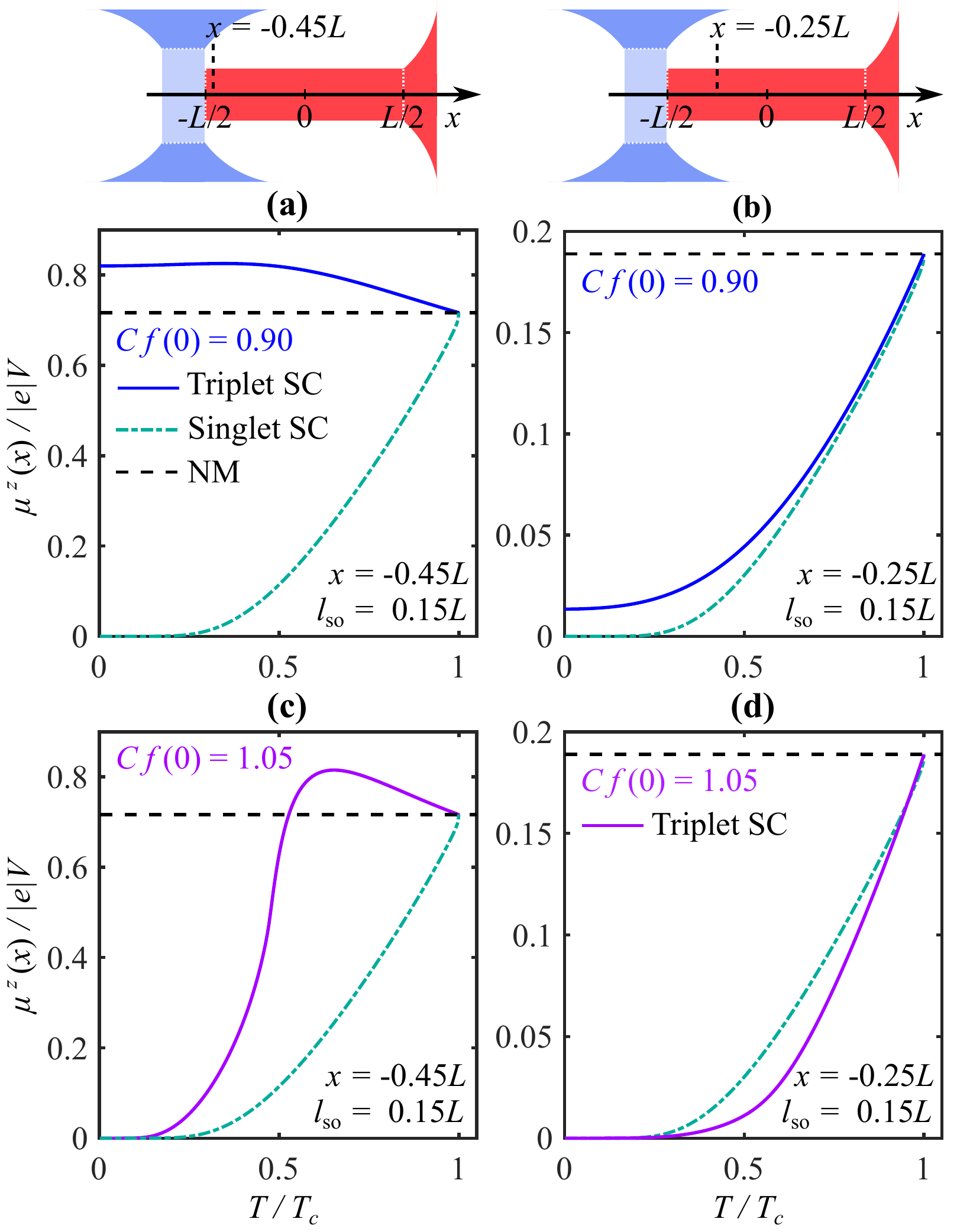}
    \caption{The non-equilibrium spin accumulation is plotted as a function of temperature in the presence of spin-orbit scattering for $l_{\text{so}}=0.15L$. Panel (a) and (b) correspond to a gapless DOS where $0\leq Cf(T) \leq0.90$ for all temperatures. Panel (c) and (d) correspond to a DOS that is gapped at low temperatures ($1<Cf(T) <1.05$), and gapless at higher temperatures ($0\leq Cf(T)\leq1$). Panel (a) and (c) are measured at a distance $0.05L$ away from the normal-metal contact, while panel (b) and (d) are measured at a distance $0.25L$ away from the normal-metal contact. The applied spin voltage is $|e|V=0.1\Delta_0$ for all panels. The triplet pairing follows the model described by Eq.~\ref{eq:type1}, however the model described by Eq.~\ref{eq:type2} gives similar results.}
    \label{fig:temperature_so}
\end{figure}

While a conventional superconductor has a giant spin-flip relaxation for energies close to the gap edge \cite{morten_prb_04}, odd-frequency superconductivity does not renormalize the average spin-flip scattering length. In fact, we find that the roles of the spin-flip and spin-orbit scattering are opposite compared to the spin-singlet case as can be seen from Eqs.~\eqref{eq:HS} and~\eqref{eq:HT}. Qualitatively, this is reasonable since spin-flip caused by magnetic impurities does not leave the spin part $(\uparrow\downarrow - \downarrow\uparrow)$ of a conventional singlet superconductor invariant. Such spin-flip does, however, leave the spin-part $(\uparrow\downarrow + \downarrow\uparrow)$ of an $S_z=0$ triplet superconductor invariant. The spin-orbit relaxation length increases at the onset of the odd-frequency superconductivity, and if a gap is present there is a giant renormalization of the spin-orbit relaxation length for energies close to the gap edge.
This is demonstrated in Fig.~\ref{fig:DOS_and_position}(c) and~(d). 
In panel~(c), only spin-flip relaxation is present, and the non-equilibrium spin-accumulation in the odd-frequency superconductor relaxes at the same rate as in the normal-metal. In the conventional superconductor, the spins relax more rapidly.
In panel~(d), only spin-orbit relaxation is present, and the non-equilibrium spin accumulation relaxes rapidly inside the odd-frequency superconductor. It relaxes at the same rate for the normal-metal and inside the conventional superconductor.
In this case, the non-equilibrium spin accumulation crosses from above to below that of the normal state, meaning that the spin-accumulation will behave qualitatively different depending on at which position it is measured.

\subsection{Temperature dependence}

Finally, we study the temperature dependence of the non-equilibrium spin accumulation.
In Fig.~\ref{fig:temperature_sf}, we consider the non-equilibrium spin accumulation for $Cf(0)=0.90$ (panel (a)) and $Cf(0)=1.05$ (panel (b)) in the presence of spin-flip scattering.
When $Cf(T)<1$ at zero temperature, the DOS is gapless for all temperatures up to $T_c$. The high number of available states causes the spin injection at the normal-metal contact to be higher than in the normal state, and the spin-flip scattering rate is the same. Therefore, the non-equilibrium spin accumulation will stay larger than in the normal state for all temperatures regardless of at which position we choose to measure it. This is demonstrated in Fig.~\ref{fig:temperature_sf}(a). In comparison, the spin-accumulation in a conventional superconductor relaxes quickly as the temperature is decreased \cite{morten_prb_04}.
When $Cf(T)>1$ at $T=0$, the DOS goes through a transition from gapped to gapless as the temperature increases. 
In the absence of spin-orbit relaxation, this causes a sharp peak in the spin accumulation at the temperature where this transition happens, as shown in Fig.~\ref{fig:temperature_sf}(b). For lower temperatures, the spin accumulation decreases as the gap widens, leaving fewer available states.

In Fig.~\ref{fig:temperature_so}, we consider the non-equilibrium spin accumulation for the same values of $Cf(T)$ in the presence of spin-orbit scattering. In this case, the spins relax quickly inside the odd-frequency superconductor. Thus, when the superconductor is gapless for all temperatures ($Cf(T)=0.90$), the spin accumulation can either be larger than in the normal state as shown in Fig.~\ref{fig:temperature_so}(a) or smaller as shown in Fig~\ref{fig:temperature_so}(b) depending on how close to the normal-metal contact it is measured.
In the case where the superconductor transitions from gapped to gapless as the temperature increases, remnants of the peak seen in the absence of spin-orbit scattering (Fig.~\ref{fig:temperature_sf}(b)) only appear close to the normal-metal contact as can be seen in Fig.~\ref{fig:temperature_so}(c). Further away, spin-orbit relaxation causes the spin accumulation to be even smaller than for conventional superconductors.

Note that in Figs.~\ref{fig:temperature_sf} and~\ref{fig:temperature_so}, we have chosen a small spin voltage in order to minimize the error from leaving out the local minimum in the DOS (see Sec.~\ref{s:DOS}). However, close to $T_c$ this will inevitably cause an overestimation in the spin injection from the normal-metal contact. This should however not cause any qualitative changes, since the spin accumulation $\mu ^z$ has to reach its normal-state value at $T_c$. It would rather cause a small reduction in the spin accumulation for temperatures close to $T_c$, and some smoothing of the peak in Fig.~\ref{fig:temperature_sf}(b).

\section{Outlook and concluding remarks}

In this work, we have shown how the non-equilibrium spin accumulation in an odd-frequency superconductor differs qualitatively in several ways from that of a conventional superconductor. First, the density of states of the odd-frequency superconductor allows for a spin injection that is larger than in the normal state. Moreover, it can transition from gapless to gapped as the temperature is decreased, causing a peak in the spin injection at a certain transition temperature below the critical temperature $T_c$ of the superconductor.
Secondly, the roles of the spin-flip and spin-orbit impurity scattering are interchanged compared to what is the case for conventional superconductors. This causes a rapid spin-orbit relaxation, and robustness in the presence of magnetic impurities.

There are several additional interesting effects which can be implemented by adjusting the setup presented in Fig.~\ref{fig:model}.
First, when injecting a spin-polarized charge current directly into a superconductor, it has been shown that the spin injection increases compared to the normal-state since the opening of the gap causes a large spin-splitting as spins accumulate at the interface \cite{jedema_prb_99,takahashi_prb_03,poli_prl_08}. At low temperatures where the density of states is gapped, this should also occur for the odd-frequency superconductors considered here. 
Secondly, we have considered the injected spin to be carried entirely by quasi-particles. By allowing the Cooper pairs to be spin-polarized with respect to the polarization axis of the injected spins, we open for the possibility that Cooper pairs contribute to the spin transport. Spins can then be injected at energies below the gap edge. 
However, this can only be described by allowing the retarded Green's function to deviate from its equilibrium value and calls for a self-consistent solution for the superconducting pairing.
Third, we have considered an effective odd-frequency superconductor in the absence of spin-splitting fields. Externally applied magnetic fields are known to enhance the spin accumulation inside conventional superconductors \cite{yang_nm_10,hubler_prl_12,quay_np_13,wolf_prb_13}. In the absence of spin splitting, spin accumulation is a direct consequence of imbalance in the spin distribution function, while in the presence of spin-splitting the imbalance in the energy distribution function gives an additional contribution due to coupling between the different modes \cite{silaev_prl_15}. A similar coupling is expected to take place for odd-frequency superconductors.
Finally, we have considered transparent boundaries between the metallic contact and the superconductor. 
In reality, there would be some contact resistance restricting the spin injection through the interface. However, a comparison of the spin accumulation above and below the superconducting critical temperature should yield qualitatively the same result although the overall signal is weaker.
In experiments, the advantage of using a tunnel barrier rather than a metallic contact is that it minimizes the proximity effect between the contact and the superconductor, which leads to a suppression in the superconducting gap close to the interface.

\begin{acknowledgements}

This work was supported by the Research Council of Norway through its Centres of Excellence funding scheme grant 262633 QuSpin.

\end{acknowledgements}

\begin{appendix}

\section{Derivation of the Usadel equation}
\label{appendix:Usadel}

We here give some more details about the underlying assumptions of the Usadel equation given in Eq.~\eqref{eq:usadel}.
Our starting point for deriving the Usadel equation is the continuum Hamiltonian
\begin{align}
    H(\ver,t)&=\int d\ver\: \sum_{\sigma}\psi_{\sigma}^{\dagger} (\ver,t)\Big(-\frac{1}{2m}\nabla_{\ver}^2 -\mu\Big)\psi_{\sigma}(\ver,t)\notag\\
    &+\frac{1}{2}\int d\ver \: \big[\Delta(\ver)\psi^{\dagger}_{\uparrow}(\ver,t)\psi_{\downarrow}^{\dagger}(\ver,t)+\text{h.c.}\big]\notag\\
    &+\int d\ver\: \sum_{\sigma,\sigma'}\psi_{\sigma}^{\dagger} (\ver,t)U_{\sigma,\sigma'}^{\text{tot}}(\ver)\psi_{\sigma'}(\ver,t),
\label{eq:hamiltonian}
\end{align}
where $\psi_{\sigma}^{(\dagger)}(\ver,t)$ is a field operator annihilating (creating) a spin-$\sigma$ electron at position $\ver$ and time $t$. 
The first term introduces the kinetic energy for electrons of mass $m$, and the chemical potential $\mu$.
The second term describes superconducting attractive interaction in the mean field approximation. The superconducting gap is defined as $\Delta(\ver)= V\left<\psi_{\uparrow}(\ver)\psi_{\downarrow}(\ver)\right>$. 
The last term introduces the total scattering potential from the impurities.

We define a four-vector field operator in Nambu~$\otimes$~spin space as
\be
    \hat{\psi}(\ver,t)=[\psi_{\uparrow}(\ver,t)\hspace{0.5em}\psi_{\downarrow}(\ver,t)\hspace{0.5em}\psi^{\dagger}_{\uparrow}(\ver,t)\hspace{0.5em}\psi^{\dagger}_{\downarrow}(\ver,t)]^T .
\ee
We also define the retarded, advanced and Keldysh Green's functions in Nambu~$\otimes$~spin space as
\begin{align}
    [\hat{G}^R (1,2)]_{i,j}
    =&-i\Theta(t_1 -t_2 )\notag\\
    &\sum_k (\hrhot)_{ik}\big<\big\{[{\psi}(1)]_k , [\hat{\psi}^{\dagger} (2)]_j \big\}\big>,~\label{eq:def_GR}\\
    [\hat{G}^A (1,2)]_{i,j} 
    =&\phantom{+}i\Theta(t_2 -t_1 )\notag\\
    &\sum_k (\hrhot)_{ik}\big<\big\{[\hat{\psi}(1)]_k , [\hat{\psi}^{\dagger} (2)]_j \big\}\big>,\label{eq:def_GA}\\
    [\hat{G}^K (1,2)]_{i,j} 
    =&-i\sum_k (\hrhot)_{ik}\big<\big[[\hat{\psi}(1)]_k , [\hat{\psi}^{\dagger} (2)]_j \big]\big>,\label{eq:def_GK}
\end{align}
respectively,
where $(1,2)$ is short-hand notation for $(\ver_1 ,t_1 ,\ver_2 ,t_2 )$. 
These are elements of the Green's function $\check{G}(1,2)$ in Keldysh space as defined for the quasi-classical Green's function in Eq.~\eqref{eq:greens_function_keldysh_space}. 
From the Heisenberg equations of motion for the field operators, we find that the equations of motion for the Keldysh space Green's function can be written as
\begin{align}
    [i\partial_{t_1}\hrhot -\hat{H}(\ver_1)]\check{G}(1,2)&=\delta(1-2)\check{\rho}_0,\label{eq:kinetic_eq_1}\\
    \check{G}(1,2)[i\partial_{t_2}\hrhot -\hrhot\hat{H}(\ver_2 )\hrhot]^{\dagger}&=\delta(1-2)\check{\rho}_0 .\label{eq:kinetic_eq_2}
\end{align}
where
\begin{align}
    \hat{H}(\ver)&=\left(-\frac{1}{2m}\nabla_{\ver}^2-\mu\right)\hrhoz
    -\hat{\Delta}^{\text{S}}(\ver)+\hat{U}_{\text{tot}}(\ver).
\end{align}
The scattering potential matrix $\hat{U}_{\text{tot}}(\ver)=U(\ver)+\hat{U}_{\text{so}}(\ver)+\hat{U}_{\text{sf}}(\ver)$ describe scattering on non-magnetic impurities, spin-orbit impurity scattering, and scattering on magnetic impurities, respectively.
The scattering potentials are given by
\begin{align}
U(\ve{r})&=\sum_i u(\ve{r}-\ve{r}_i ),\\
\hat{U}_{\text{so}}(\ver)&=\sum_i i\alpha [\hat{\rho}_3 \hat{\ve{\sigma}}\times\nabla_{\ve{r}}u(\ve{r}-\ve{r}_i )]\cdot\nabla_{\ve{r}},\\
\hat{U}_{\text{sf}}(\ver)&=\sum_i u_{\text{m}} (\ver-\ver_i )\vehsigma\cdot \ve{S}_i,
\end{align}
where $u(\ve{r}-\ve{r}_i )$ and $u_{\text{m}} (\ve{r}-\ve{r}_i )$ are the scattering potentials of a single non-magnetic and magnetic impurity, and $\ve{S}_i$ is the spin of the magnetic impurity at position $\ve{r}_i$.

In order to solve Eqs.~\eqref{eq:kinetic_eq_1} and~\eqref{eq:kinetic_eq_2}, we must replace the impurity potentials by self energies. To do this, we split the Hamiltonian up into two parts, $\hat{H}(\ver)=\hat{H}_0 (\ver) +\hat{U}_{\text{tot}}(\ver)$, where $\hat{H}_0 (\ver)$ describes the system in the absence of impurity scattering. We introduce self-energies through the Dyson equations
\begin{align}
\check{G}(1,2)&=\check{G}_0 (1,2)+\check{G}_0\bullet\hat{\Sigma}\bullet\check{G}(1,2),\\
\check{G}(1,2)&=\check{G}_0 (1,2)+\check{G}\bullet\hat{\Sigma}^{\dagger }\bullet\check{G}_0 (1,2),
\end{align}
where the self-energies are defined as $\hat{\Sigma}(1,2)=\delta(1-2)\hat{U}_{\text{tot}}(\ver_2 )$.
Above, $\check{G}_0 (1,2)$ is the Green's function in the absence of impurity scattering, and we have introduced the bullet product
\be
A\bullet B(1,2)=\int d3\: A(1,3)B(3,2).
\ee
We solve the Dyson equations iteratively within the self-consistent Born approximation by neglecting terms above the second order in $\hat{\Sigma}\bullet\check{G}$ and $\check{G}\bullet\hat{\Sigma}^{\dagger}$. Since we are not interested in one specific impurity configuration, we take the average over all impurities, 
\be
\Big<\hdots\Big>_{\text{av}}=\prod_{n=1}^N \left(\frac{1}{\mathscr{V}}\int d\ver_n \: \right)\hdots,
\ee
where $\mathscr{V}$ is the volume of the system. We assume that the Green's function is approximately equal to its impurity-averaged value.
By acting with $[i\partial_{t_1}\hrhot -\hat{H}_0 (\ver_1 )]$ and $[i\partial_{t_2}\hrhot -\hrhot\hat{H}_0 (\ver_2 )\hrhot]$ on the resulting equations, we obtain expressions on a similar form as Eqs.~\eqref{eq:kinetic_eq_1} and~\eqref{eq:kinetic_eq_2} where the impurity potentials are replaced by expressions involving self-energies and impurity averaged Green's functions. Subtracting the two equations, we find that
\begin{align}
    &[i\partial_{t_1}\hrhot -\hat{H}_0 (\ver_1 )]\check{G}_{\text{av}}(1,2)\notag\\
    &-\check{G}_{\text{av}}(1,2)[i\partial_{t_2}\hrhot -\hrhot\hat{H}_0 (\ver_2 )\hrhot]^{\dagger}\notag\\
   &-[\left<\hat{\Sigma}\bullet\check{G}_{\text{av}}\bullet\hat{\Sigma}\right>_{\text{av}}\combullet\check{G}_{\text{av}} ](1,2)=0.\label{eq:equations_averaged}
\end{align}
In order to arrive at Eq.~\eqref{eq:usadel} we now need to introduce several approximations to the above equation.

We first introduce center-of-mass and relative coordinates $\veR=(\ver_1 +\ver_2 )/2$ and $\ver=\ver_1 -\ver_2$, as well as absolute and relative time coordinates $T=(t_1 +t_2 )/2$ and $t=t_1 - t_2 $. We assume that the Green's function is independent of the absolute time coordinate, and that all quantities varies slowly in space compared to the Fermi wavelength. This allows us to keep only the first order gradients in the center of mass coordinate.
We introduce the Fourier transform and its inverse, 
\begin{align}
\check{G}_{\text{av}}(\veR,\ve{p},\epsilon)&=\int d\ve{r}\int dt \: e^{-i\ve{p}\cdot\ve{r}+i\epsilon t}\check{G}_{\text{av}}(\veR,\ve{r},t) ,\label{eq:FT}\\
\check{G}_{\text{av}}(\veR,\ve{r},t)&=\int \frac{d\ve{p}}{(2\pi)^3}\int \frac{d\epsilon}{2\pi} e^{i\ve{p}\cdot\ve{r}-i\epsilon t} \check{G}_{\text{av}}(\veR,\ve{p},\epsilon) .\label{eq:iFT}
\end{align}
Under these assumptions, the Fourier transform of the bullet product between two functions $A(\veR,\vep,\epsilon)$ and $B(\veR,\vep,\epsilon)$ is given by
\begin{align}
A\bullet B(\veR,\vep,\epsilon)=&A(\veR,\vep,\epsilon) B(\veR,\vep,\epsilon)\notag\\ &+\frac{i}{2}[\nabla_{\veR}A(\veR,\vep,\epsilon)\cdot\nabla_{\vep}B(\veR,\vep,\epsilon)\notag\\
&-\nabla_{\vep}A(\veR,\vep,\epsilon)\cdot\nabla_{\veR}B(\veR,\vep,\epsilon)].
\label{eq:gradient_approximation}
\end{align}

Next, we assume that the absolute value of the momentum $p$ is approximately equal to the Fermi momentum $p_{\text{F}}$. This allows us to apply the quasi-classical approximation
\be
\int\frac{d\ve{p}}{(2\pi)^3}\:\check{G}_{\text{av}}(\veR,\ve{p},\epsilon)\approx N_0 \int d\xi_{p_{\text{F}}} \:\int \frac{d\ve{e}_{p_{\text{F}}}}{4\pi}\:\check{G}_{\text{av}}(\veR, \ve{p}_{\text{F}},\epsilon).\label{eq:quasi-cl_approx}
\ee
Above, $N_0$ is the DOS at the Fermi level, $\xi_{p_{\text{F}}}=p_{\text{F}}^2/2m$, and $\ve{e}_{p_{\text{F}}}=\ve{p}_{\text{F}}/p_{\text{F}}$ describes the direction of the momentum.
We will use the short-hand notation
$\left<\hdots\right>_{p_{\text{F}}}=\int (d\ve{e}_{p_{\text{F}}}/4\pi)\:$ for the average over all directions of the momentum.
Moreover, we introduce the quasi-classical Green's function
\be
\check{g}_{\text{av}}(\ve{R},\ve{p}_{\text{F}} ,\epsilon )= \frac{i}{\pi}\int d\xi_{p_{\text{F}}} \: \check{G}_{\text{av}}(\ve{R},\ve{p}_{\text{F}},\epsilon).\label{eq:quasi-cl_GF}
\ee
In the diffusive limit, the quasi-classical Green's function can be approximated as
\be
\check{g}_{\text{av}}(\ve{R},\ve{p}_{\text{F}} ,\epsilon)\approx\check{g}_{\text{av}}^{\text{s}} (\ve{R},\epsilon)+\ve{e}_{p_{\text{F}}}\cdot\check{\ve{g}}_{\text{av}}^{\text{p}} (\ve{R},\epsilon).\label{eq:g->gs+gp}
\ee
We assume that $|\check{\ve{g}}_{\text{av}}^{\text{p}}(\veR,\epsilon) |\ll \check{g}_{\text{av}}^{\text{s}}(\veR,\epsilon)$ and neglect terms of second order in $\check{\ve{g}}_{\text{av}}^{\text{p}}(\veR,\epsilon)$.

After applying all these approximations to Eq.~\eqref{eq:equations_averaged}, we separate out the even contributions in $\vece_{p_{\text{F}}}$ by averaging over all $\vece_{p_{\text{F}}}$. 
We next separate out the odd contributions in $\vece_{p_{\text{F}}}$ by multiplying the equation by $\vece_{p_{\text{F}}}$ before doing the averaging.
In the odd equation, we assume that the scattering on non-magnetic impurities dominates over all other terms, and use the normalization condition
\begin{align}
    \check{g}_{\text{av}}(\veR,p_{\text{F}},\epsilon)\check{g}_{\text{av}}(\veR,p_{\text{F}},\epsilon)=\check{\rho}_0 
\end{align}
to express $\check{\ve{g}}_{\text{av}}^{\text{p}}(\veR,\epsilon)$ in terms of $\check{g}_{\text{av}}^{\text{s}}(\veR,\epsilon)$ as 
\begin{align}
    \check{\ve{g}}_{\text{av}}^{\text{p}}(\veR,\epsilon)= -\tau v_{\text{F}} \check{g}_{\text{av}}^{\text{s}}(\veR,\epsilon)\nabla_{\veR}\check{g}_{\text{av}}^{\text{s}}(\veR,\epsilon).
    \label{eq:gp_expression}
\end{align}
This leaves us with contributions only from second order terms in each of the three scattering potentials. Cross terms including two different types of scattering potential either disappear when we neglect terms from the odd equation, or they are neglected due to averaging over all directions of the spins of the magnetic impurities. In treating second order terms in the magnetic impurity potential, the same averaging over spin directions causes cross terms between two different magnetic impurities to give zero contribution. We can then write $S_{i}S_{j}=S(S+1)\delta_{i,j}$.
Inserting Eq.~\eqref{eq:gp_expression} into the even equation results in the Usadel equation given in Eq.~\eqref{eq:usadel}.

\end{appendix}

\end{document}